\documentclass[preprint,showpacs]{revtex4-1}
\usepackage{amsmath,amssymb,amsfonts,dcolumn,color,graphicx,graphics,latexsym,placeins,epsfig,tikz}
\usetikzlibrary{arrows,shapes}
\usepackage{subfigure,rotating,hyperref,bm}

\newcommand{\be}{\begin{equation}}
\newcommand{\ee}{\end{equation}}
\newcommand{\ba}{\begin{eqnarray}}
\newcommand{\ea}{\end{eqnarray}}

\begin{document}

\title{\Large \bf Lorentzian wormholes in Eddington-inspired Born-Infeld gravity}

\author{Rajibul Shaikh}
\email{rajibulshaikh@cts.iitkgp.ernet.in}
\affiliation{Centre for Theoretical Studies, Indian Institute of Technology Kharagpur,\\ Kharagpur 721 302, India.}

\begin{abstract}
\noindent We show that it is possible to construct a wide class of Lorentzian wormholes in Eddington-inspired Born-Infeld gravity with a stress energy which does not violate the weak or null energy condition. The wormholes exist in a certain region of the parameter space. In fact, it is shown that there is a critical value of a parameter defined in our work, below which we have wormholes. Above the critical value, we have a regular black hole spacetime. We put a restriction on the equation of state parameter $\alpha$ ($p_{\theta}=\alpha \rho$) to have wormholes. We also put a lower limit on both the theory parameter $|\kappa|$ and the throat radius, to restrict the tidal acceleration (at the throat) below one Earth gravity. As a special case of our general solution, we retrieve the wormhole supported by an electric field for a charge-to-mass ratio greater than the critical value $\left(\frac{Q}{M}\right)_c\approx 1.144$.
\end{abstract}


\maketitle

\section{Introduction}
\noindent A wormhole is a topological or geometrical structure that constitutes a short cut or tunnel between two universes or two separate regions of the same universe \cite{morris1,morris2,visser1}. The existence of such structures implies the violation of the convergence condition, as required by the Raychaudhuri equation \cite{hawking,wald}. In general relativity (GR), a violation of the convergence condition implies a violation of various energy conditions (weak, strong, null, dominant, etc.) \cite{hawking,wald}. Therefore, in the context of GR, one needs energy-condition-violating matter (often called ``exotic matter") to sustain a wormhole \cite{morris1,morris2,visser1}. However, in some alternative or modified theories of gravity, the violation of the convergence condition may not imply a violation of the energy conditions \cite{capozziello1,capozziello2}. In such theories, these two conditions are separate. Therefore, in such theories, one can have a wormhole geometry with the matter satisfying all the energy conditions but violating the convergence condition.

Born-Infeld  inspired gravity is a modified gravity theory first suggested by Deser and Gibbons \cite{deser}, inspired by the earlier work of Eddington \cite{eddington} and the nonlinear electrodynamics of Born and Infeld \cite{born}. They considered the metric formulation of the Born-Infeld gravitational action. Later, Vollick \cite{vollick1,vollick2,vollick3} considered the Born-Infeld gravity action in the Palatini formalism and introduced the matter fields in a nonconventional way. Recently, Banados and Ferreira \cite{banados} considered the matter coupling in a different way, simpler than that introduced in Vollick's work. We shall focus on the Banados and Ferreira formulation of Born-Infeld gravity \cite{banados}, commonly cited as Eddington-inspired Born-Infeld gravity (EiBI). Various aspects of EiBI gravity have been studied by many researchers in the recent past. Spherically symmetric solutions, cosmology, and astrophysical aspects have been worked out \cite{EiBIcharged1,EiBIcharged2,EiBIcharged3,EiBIsol1,EiBIsol2,EiBIsol3,EiBIsol4,
EiBIsol5,EiBIsol6,EiBIsol7,EiBIsol8}. The bouncing solutions obtained in EiBI cosmology, with matter satisfying all the energy conditions, implies a violation of the convergence condition. This gives a hint of having a wormhole geometry supported by ordinary matter, in EiBI gravity. Attempts have been made to construct a wormhole geometry in this modified gravity \cite{EiBIsol7,EiBIwormhole1,EiBIwormhole2}. The authors in \cite{EiBIsol7} obtained a wormhole geometry supported by pressureless dust, in three dimensions. However, the energy density is made positive with the help of the negative cosmological constant. Without the cosmological constant, the energy density is negative on one side of the throat and positive on the other side. In \cite{EiBIwormhole1}, the authors consider an anisotropic fluid and obtain an exact solution for a wormhole geometry. However, in their case, the energy density is negative and, hence, the energy conditions are violated. By writing the electrically charged solution in EiBI gravity in a different gauge and noting the bounce of the radial function, the authors in \cite{EiBIwormhole2} show that, for negative Eddington-Born-Infeld theory parameter $\kappa$, this solution signals the presence of a wormhole. In this paper, we try to construct a wormhole supported by an anisotropic fluid, satisfying all the energy conditions. As a special case of our general solution, we obtain a wormhole geometry supported by an electric field as mentioned in \cite{EiBIwormhole2}.

\section{Eddington-inspired Born-Infeld gravity and the spacetime metric}
\noindent The action in EiBI gravity is given by
\begin{eqnarray}
S_{BI}[g,\Gamma,\Psi]&=&\frac{c^4}{8\pi G\kappa}\int d^4x\left[\sqrt{-\left\vert g_{\mu\nu}+\kappa R_{\mu\nu}(\Gamma)\right\vert}-\lambda \sqrt{-g}\right]+S_{M}(g,\Psi),
\nonumber
\end{eqnarray}
where $c$ is the speed of light, $G$ is Newton's gravitational constant, $\lambda=1+\kappa\Lambda$, $R_{\mu\nu}(\Gamma)$ is the symmetric part of the Ricci tensor built with the connection $\Gamma$ and $S_{M}(g,\Psi)$ is the action for the matter field. $\Lambda$ is the cosmological constant. Variations of this action with respect to the metric tensor $g_{\mu\nu}$ and  the connection $\Gamma$ yield, respectively, \cite{banados,EiBIsol6,EiBIwormhole2}
\begin{equation}
\sqrt{-q}q^{\mu\nu}=\lambda \sqrt{-g}g^{\mu\nu}-\bar{\kappa} \sqrt{-g}T^{\mu\nu}
\label{eq:field_equation1}
\end{equation}
\begin{equation}
\nabla^\Gamma_\alpha \left(\sqrt{-q} q^{\mu\nu}  \right)=0,
\label{eq:metric_compatibility}
\end{equation}
where $\bar{\kappa}=\frac{8\pi G\kappa}{c^4}$, $\nabla^\Gamma$ denotes the covariant derivative defined by the connection $\Gamma$ and $q^{\mu\nu}$ is the inverse of the auxiliary metric $q_{\mu\nu}$ defined by
\begin{equation}
q_{\mu\nu}=g_{\mu\nu}+\kappa R_{\mu\nu}(\Gamma).
\label{eq:field_equation2}
\end{equation}
To obtain these equations, it is assumed that both the connection $\Gamma$ and the Ricci tensor $R_{\mu\nu}(\Gamma)$ are symmetric, i.e., $\Gamma^\mu_{\nu\rho}=\Gamma^\mu_{\rho\nu}$ and $R_{\mu\nu}(\Gamma)=R_{\nu\mu}(\Gamma)$. Equation (\ref{eq:metric_compatibility}) gives the metric compatibility equation which yields
\begin{equation}
\Gamma^\mu_{\nu\rho}=\frac{1}{2}q^{\mu\sigma}\left(q_{\nu\sigma,\rho}+q_{\rho\sigma,\nu}-q_{\nu\rho,\sigma} \right).
\nonumber
\end{equation}
Therefore, the connection $\Gamma^\mu_{\nu\rho}$ is the Levi-Civita connection of the auxiliary metric $q_{\mu\nu}$.

For the matter part, we consider an anisotropic fluid having an energy-momentum tensor of the form
\begin{equation}
T^\mu_{\;\nu}=diag(-\rho,p_r,p_\theta,p_\theta)=diag(-\rho,-\rho,\alpha\rho,\alpha\rho).
\nonumber
\end{equation}
A similar type of energy momentum have been considered to obtain static, spherically symmetric, exact solutions in Einstein gravity coupled to quintessential matter \cite{quintessence}. To satisfy all the energy conditions, we must have $\rho \geq 0$ and $0 \leq \alpha \leq 1$. For $\alpha=1$, it can represent the energy-momentum tensor of a Maxwell field, given by
\begin{equation}
T_{\mu\nu}=\frac{1}{4\pi}\left(F_{\mu\rho}F^{\rho}_{\;\nu}-\frac{1}{4}g_{\mu\nu}F_{\rho\sigma}F^{\rho\sigma}\right).
\nonumber
\end{equation}
Following \cite{EiBIcharged1,EiBIcharged2}, we consider the following $Ans\ddot{a}tze$ for the physical ($g$-) and auxiliary ($q$-) line element:
\begin{equation}
ds_g^2=-\psi^2(r)f(r)dt^2+\frac{dr^2}{f(r)}+r^2(d\theta ^2+\sin ^2\theta d\phi ^2)
\label{eq:physical_metric}
\end{equation}
\begin{equation}
ds_q^2=-G^2(r)F(r)dt^2+\frac{dr^2}{F(r)}+H^2(r)(d\theta ^2+\sin ^2\theta d\phi ^2).
\label{eq:auxiliary_metric}
\end{equation}
In the above $Ans\ddot{a}tze$, a gauge has been chosen so that the radial function of the $g$ metric is fixed to $r$, giving $4\pi r^2$ as the area of the two-sphere of constant $r$. As shown in \cite{EiBIsol6}, the energy momentum satisfies the conservation law with respect to the physical ($g$-) metric, i.e., $\nabla_\mu^{(g)}T^{\mu\nu}=0$, where $\nabla^{(g)}$ denotes the covariant derivative defined by the
Christoffel symbol based on $g_{\mu\nu}$. The conservation law is a consequence of assuming minimal coupling. For the metric in (\ref{eq:physical_metric}), the energy conservation equation becomes
\begin{equation}
p_r'+\frac{2}{r}\left(p_r-p_\theta \right)+\left(\rho + p_r\right)\left(\log\left[\psi(r)\sqrt{f(r)}\right]\right)'=0,
\end{equation}
where prime ($'$) denotes derivative with respect to $r$. For the energy-momentum $T^\mu_{\;\nu}=diag(-\rho,-\rho,\alpha\rho,\alpha\rho)$, integration of the above equation yields the energy density
\begin{equation}
\rho=\frac{C_0}{r^{2(\alpha+1)}},
\end{equation}
where $C_0$ is an integration constant with dimension dependent on the value of $\alpha$. For $\lambda=1$ (i.e., $\Lambda=0$), Eq. (\ref{eq:field_equation1}) gives the following relations between the physical and auxiliary metric:
\begin{equation}
f(r)=F(r)(1-\bar{\kappa}\alpha\rho), \hspace{0.5cm} \psi(r)=G(r)(1-\bar{\kappa}\alpha\rho)^{-1}
\nonumber
\end{equation}
\begin{equation}
H(r)=r\sqrt{1+\bar{\kappa}\rho}.
\nonumber
\end{equation}
Using the above-mentioned $Ans\ddot{a}tze$ in (\ref{eq:field_equation2}), we obtain
\begin{equation}
\frac{G''}{G}+\frac{F''}{2F}+\frac{3G'F'}{2GF}+\frac{H'F'}{HF}+\frac{2H'G'}{HG}=\frac{1}{\kappa F}\left[\frac{1}{1-\bar{\kappa}\alpha\rho}-1\right]
\label{eq:fieldeqn1}
\end{equation}
\begin{equation}
\frac{G''}{G}+\frac{F''}{2F}+\frac{3G'F'}{2GF}+\frac{H'F'}{HF}+\frac{2H''}{H}=\frac{1}{\kappa F}\left[\frac{1}{1-\bar{\kappa}\alpha\rho}-1\right]
\label{eq:fieldeqn2}
\end{equation}
\begin{equation}
\frac{H''}{H}+\frac{H'^2}{H^2}+\frac{H'F'}{HF}+\frac{H'G'}{HG}-\frac{1}{H^2F}=\frac{1}{\kappa F}\left[\frac{1}{1+\bar{\kappa}\rho}-1\right].
\label{eq:fieldeqn3}
\end{equation}
Comparing Eqs. (\ref{eq:fieldeqn1}) and (\ref{eq:fieldeqn2}), we get
\begin{equation}
G(r)=H'(r)=\left[1-\frac{\bar{\kappa}\alpha C_0}{r^{2(\alpha+1)}}\right]\left[1+\frac{\bar{\kappa} C_0}{r^{2(\alpha+1)}}\right]^{-\frac{1}{2}},
\nonumber
\end{equation}
which gives
\begin{equation}
\psi(r)=\left[1+\frac{\bar{\kappa} C_0}{r^{2(\alpha+1)}}\right]^{-\frac{1}{2}}.
\nonumber
\end{equation}
We are now left with Eqns. (\ref{eq:fieldeqn2}) and (\ref{eq:fieldeqn3}), and one unknown function $F(r)$. One of these equations can be integrated to obtain $F(r)$. The energy conservation equation which we have already considered, forces the other equation to be satisfied. Solving Eq. (\ref{eq:fieldeqn3}), we get
\begin{eqnarray}
F&=& \frac{1}{HH'^2}\left[C_1+H+\frac{1}{\kappa}\int H^2H'\left(\frac{1}{1+\frac{\bar{\kappa} C_0}{r^{2(\alpha+1)}}}-1\right) dr\right] \nonumber \\
&=& \frac{1}{HH'^2}\left[C_1+H-\frac{H^3}{3\kappa}+\frac{1}{\kappa}\int \frac{H^2H'}{1+\frac{\bar{\kappa} C_0}{r^{2(\alpha+1)}}} dr\right],
\label{eq:integration}
\end{eqnarray}
where $C_1$ is another integration constant. Therefore, we obtain (see Appendix)
\begin{eqnarray}
f(r)&=&\frac{1+\frac{\bar{\kappa} C_0}{r^{2(\alpha+1)}}}{1-\frac{\bar{\kappa}\alpha C_0}{r^{2(\alpha+1)}}}\left[1-\frac{\bar{\kappa}C_0}{3\kappa r^{2\alpha}}+\frac{C_1}{r\sqrt{1+\frac{\bar{\kappa} C_0}{r^{2(\alpha+1)}}}}-\frac{2(\alpha+1)\bar{\kappa} C_0}{3\kappa r\sqrt{1+\frac{\bar{\kappa} C_0}{r^{2(\alpha+1)}}}}\int \frac{dr}{r^{2\alpha}\sqrt{1+\frac{\bar{\kappa} C_0}{r^{2(\alpha+1)}}}}\right].
\end{eqnarray}
To fix up the integration constant $C_1$, we take the vacuum limit. In vacuum ($\rho=0$, i.e., $C_0=0$), we must recover the Schwarzschild solution. This gives $C_1=-\frac{2GM}{c^2}$, $M$ being related to the mass.

\section{Constructing The Wormhole}
\noindent The existence of a wormhole spacetime requires the presence of a minimal surface called the throat. The minimality of the throat is reinterpreted as divergent null rays passing through the throat. This requires the violation of the null convergence condition, from the Raychaudhuri equation. The Raychaudhuri equation for a bundle of light rays with vanishing shear and rotation is given by
\begin{equation}
\frac{d\hat{\theta}}{d\lambda}+\frac{1}{2}\hat{\theta}^2+R_{\alpha\beta}\hat{u}^\alpha \hat{u}^\beta=0,
\nonumber
\end{equation}
where $\hat{u}^\alpha$ is the four velocity of the light ray. For a radial null ray traveling through the wormhole in the equatorial plane ($\theta=\pi/2$), $\hat{u}^t=1/(\psi^2f)$ and $\hat{u}^r=\pm 1/\psi$. Therefore, $\hat{\theta}$, $d\hat{\theta}/d\lambda$ and $R_{\alpha\beta}\hat{u}^\alpha \hat{u}^\beta$ are given by
\begin{equation}
\hat{\theta}=\pm \frac{2}{r}\sqrt{1+\frac{\bar{\kappa} C_0}{r^{2(\alpha+1)}}}, \hspace{0.1cm} \frac{d\hat{\theta}}{d\lambda}=-\frac{2}{r^2}\left[1+\frac{(\alpha+2)\bar{\kappa} C_0}{r^{2(\alpha+1)}}\right]
\nonumber
\end{equation}
\begin{equation}
R_{\alpha\beta}\hat{u}^\alpha \hat{u}^\beta=\frac{2(\alpha+1)\bar{\kappa} C_0}{r^{2(\alpha+2)}},
\nonumber
\end{equation}
where $-$ and $+$ signs are for ingoing and outgoing rays, respectively. To  satisfy energy conditions, we must have $C_0>0$ and $0 \leq \alpha \leq 1$. Therefore, we must have $\kappa<0$ to have a wormhole solution. It is to be noted that, for $\kappa <0$, the expansion vanishes at $r_0=(|\bar{\kappa}|C_0)^{\frac{1}{2(\alpha+1)}}$. Therefore, the throat must be at $r=r_0$, implying that the metric function $f(r)$ must vanish at $r=r_0$. For $\kappa<0$, the physical metric functions become
\begin{equation}
\psi(r)=\left[1-\frac{r_0^{2(\alpha+1)}}{r^{2(\alpha+1)}} \right]^{-\frac{1}{2}}
\nonumber
\end{equation}
\begin{eqnarray}
f(r)&=&\frac{1-\frac{r_0^{2(\alpha+1)}}{r^{2(\alpha+1)}}}{1+\alpha\frac{r_0^{2(\alpha+1)}}{r^{2(\alpha+1)}}}\left[1-\frac{r_0^{2(\alpha+1)}}{3|\kappa|r^{2\alpha}}-\frac{2\bar{M}}{r\sqrt{1-\frac{r_0^{2(\alpha+1)}}{r^{2(\alpha+1)}}}}-\frac{2(\alpha+1)r_0^{2(\alpha+1)}}{3|\kappa|r\sqrt{1-\frac{r_0^{2(\alpha+1)}}{r^{2(\alpha+1)}}}}I(r)\right],
\nonumber
\end{eqnarray}
where
\begin{equation}
I(r)=\int^r \frac{dr}{r^{2\alpha}\sqrt{1-\frac{r_0^{2(\alpha+1)}}{r^{2(\alpha+1)}}}}
\end{equation}
and $\bar{M}=\frac{GM}{c^2}$. The spacetime representing a static, spherically symmetric geometry is generically written as \cite{morris1}
\begin{equation}
ds^2=-e^{2\Phi(r)}dt^2+\frac{dr^2}{1-\frac{b(r)}{r}}+r^2(d\theta ^2+\sin ^2\theta d\phi ^2),
\label{eq:general_metric}
\end{equation}
where $\Phi(r)$ and $b(r)$ are, respectively, the redshift function and the wormhole shape function. To visualize the wormhole, one embeds the $t=$constant, $\theta=\pi/2$ section of the wormhole spacetime in background cylindrical coordinates ($z$, $r$, $\phi$) using an embedding function $z(r)$. The line element on the embedding surface can be written as
\begin{eqnarray}
ds_2^2&=&dz(r)^2+dr^2+r^2d\phi ^2 \nonumber \\
      &=&\left[1+\left(\frac{dz}{dr}\right)^2\right]dr^2+r^2 d\phi ^2.
\nonumber
\end{eqnarray}
Matching this with the $t=$constant, $\theta=\pi/2$ section of the metric in (\ref{eq:general_metric}), one obtains the embedding equation:
\begin{equation}
\frac{dz}{dr}=\pm \sqrt{\frac{b/r}{1-b/r}}.
\end{equation}
Imposing the minimality of the wormhole throat, one obtains the well-known flare-out condition at the throat:
\begin{equation}
\frac{d}{dz}\left(\frac{dr}{dz}\right)=\frac{b-b'r}{2b^2}>0
\end{equation}
For the metric in (\ref{eq:physical_metric}), $e^{2\Phi(r)}=\psi^2f$ and $(1-b/r)=f$. Therefore, in terms of $f(r)$, the flare-out condition reads $\frac{f'}{2(1-f)^2}>0$. It is easy to check that, for $\alpha>0$, $b/r\to 0$ (i.e. $f\to1$) and $\Phi\to 0$ as $r\to \infty$. Therefore, the spacetime is asymptotically flat. One of the important conditions to construct a wormhole is that the redshift function $\Phi(r)$ must be finite everywhere (from the throat to spatial infinity); i.e., the metric function $\psi^2f$ should always be nonzero, positive and finite. Due to the presence of the third and fourth terms in $f$, $\psi^2f$ diverges as $r\to r_0$. However, we remove this divergence by taking the following relation between $|\kappa|$, $M$ and $r_0$ (i.e. $C_0$):
\begin{equation}
\bar{M}=-\frac{(\alpha+1)r_0^{2(\alpha+1)}}{3|\kappa|}I(r_0).
\label{eq:constraint1}
\end{equation}
Using the l'Ho$\hat{\text{p}}$ital's rule, one can show that, as $r\to r_0$, the divergence in the third and fourth terms cancel each other giving a finite nonzero value. Hence, $\psi^2f$ is finite at $r=r_0$. One can make $\psi^2f$ nonzero and positive always by adjusting $|\kappa|$, $M$ and $C_0$. For $r>r_0$, $f$ should not vanish to make $\psi^2f$ nonvanishing (no-horizon condition). This can be achieved if the square-bracketed quantity in $f$ does not vanish at $r>r_0$. At $r=r_0$, we have, after using the l'Ho$\hat{\text{p}}$ital's rule,
\begin{equation}
f(r)\big|_{r_0}=0, \hspace{0.3cm} \psi^2(r)f(r)\big|_{r_0}=\frac{1}{\alpha+1}(1-x)
\end{equation}
\begin{equation}
\frac{f'}{2(1-f)^2}\big|_{r_0}=\frac{1}{r_0}(1-x),
\end{equation}
where $x=\frac{r_0^2}{|\kappa|}$. Therefore, to satisfy the flare-out condition as well as $\psi^2f>0$ at the throat, we must have $x<1$, i.e., $r_0<|\kappa|^{1/2}$. Since, for $x<1$, $f=0$ and $f'>0$ at the throat, $f(r)$ does not have any zeroes at $r>r_0$. On the other hand, it always possesses zeroes at $r>r_0$, for $x>1$. Thus, we always have a wormhole solution for $x<1$ and a regular black hole (or a wormhole whose throat is covered by an event horizon) solution for $x>1$. The critical value $x_c=1$ distinguishes the wormhole and black hole solutions.

The condition in (\ref{eq:constraint1}) makes the spacetime regular everywhere. At $r=r_0$, the scalar invariants such as the Ricci scalar $\cal{R}$ and the Kretschmann scalar $\cal{K}$ are given by
\begin{equation}
\mathcal{R}\big|_{r_0}=-\frac{1}{r_0^2}\left[2(\alpha+1)-4\left(\frac{\alpha}{3}+1\right)x\right]
\label{eq:curvature}
\end{equation}
\begin{equation}
\mathcal{K}\big|_{r_0}=\frac{4}{r_0^4}\left[\left(\alpha^2+3\right)-\frac{4}{3}\left(\alpha^2+3\right)x+\left(\frac{4}{9}\alpha^2+2\right)x^2\right].
\label{eq:kretchman}
\end{equation}
These expressions for the invariant scalars are obtained without performing the integration in $I(r)$. However, one can perform the integration separately for $\alpha=\frac{1}{2}$ and $\alpha\neq \frac{1}{2}$. For $\kappa<0$, this gives (see Appendix)
\begin{equation}
I(r) = \left\{
  \begin{array}{lr}
    \frac{2}{3}\log\left[\left(\frac{r}{r_0}\right)^{\frac{3}{2}}+\sqrt{\left(\frac{r}{r_0}\right)^{3}-1}\right] & : \alpha =\frac{1}{2}\\
  \frac{r^{1-2\alpha}}{1-2\alpha} {}_2F_1\left[\frac{1}{2},\frac{2\alpha-1}{2\alpha+2},\frac{4\alpha+1}{2\alpha+2};\left(\frac{r_0}{r}\right)^{2\alpha+2} \right] & : \alpha \neq \frac{1}{2}
  \end{array}
\right. ,
\end{equation}
where ${}_2F_1[a,b,c;x]$ represents a hypergeometric function. Therefore, the condition (\ref{eq:constraint1}) becomes
\begin{equation}
\bar{M} = \left\{
  \begin{array}{lr}
    0 & : \alpha =\frac{1}{2}\\
  \frac{(\alpha+1)r_0^3}{3(2\alpha-1)|\kappa|} \hspace{0.1cm} {}_2F_1\left[\frac{1}{2},\frac{2\alpha-1}{2\alpha+2},\frac{4\alpha+1}{2\alpha+2};1 \right] & : \alpha \neq \frac{1}{2}
  \end{array}
\right. .
\label{eq:constraint2}
\end{equation}
Thus, in order to make the spacetime regular, the Schwarzschild mass $M$ must be negative, zero and positive for $\alpha< \frac{1}{2}$, $\alpha= \frac{1}{2}$ and $\alpha> \frac{1}{2}$, respectively. If we do not want $M$ to be negative or zero, then the condition in (\ref{eq:constraint1}) is not valid for $\alpha\leq \frac{1}{2}$. In that case, we always have a singular spacetime. Therefore, with $M>0$, a wormhole solution is not possible for $\alpha\leq \frac{1}{2}$. On the other hand, we always have a wormhole solution for $\alpha> \frac{1}{2}$ provided $x<1$ and the condition in (\ref{eq:constraint1}) is satisfied.

Let us now find the nature of the auxiliary ($q$-) metric for $\kappa<0$. In this case, the auxiliary metric functions become
\begin{equation}
G^2F=\psi^2f \left(1+\alpha\frac{r_0^{2(\alpha+1)}}{r^{2(\alpha+1)}}\right), \hspace{0.1cm} F=f \left(1+\alpha\frac{r_0^{2(\alpha+1)}}{r^{2(\alpha+1)}}\right)^{-1}
\nonumber
\end{equation}
\begin{equation}
H=r\sqrt{1-\frac{r_0^{2(\alpha+1)}}{r^{2(\alpha+1)}}}.
\nonumber
\end{equation}
The nature of the $tt$ and $rr$ components of the $q$ metric are similar to that of $g$ metric; at $r=r_0$, $F(r)$ vanishes and $G^2F$ is finite. But, the area radius $H(r)$ goes to zero at $r=r_0$. Therefore, in the case where the $g$ metric gives either a wormhole or a regular black hole spacetime, the $q$ metric gives either a naked singularity or a black hole singularity depending on the value of $x$. It should be noted that the volume ratio $\tau=\sqrt{\frac{-g}{-q}}$ diverges as $r\to r_0$. As mentioned in \cite{tau}, the divergence of $\tau$ triggers the singularity avoidance in the $g$ metric in the cosmological context. However, without the condition in (\ref{eq:constraint1}), the physical metric become singular and $\tau$ also diverges as $r\to r_0$, for $\kappa<0$. Hence, in that case, a divergence of $\tau$ as $r\to r_0$ is unable to trigger any singularity avoidance in the $g$ metric.

\section{Tidal force at the throat}
\noindent For the wormhole to be traversable, the tidal force felt by a traveler (human being) traveling through it must be within a tolerable limit. In an orthonormal basis $\{e_{\hat{0}'}, e_{\hat{1}'}, e_{\hat{2}'}, e_{\hat{3}'}\}$ of the traveler frame radially moving through the wormhole, the tidal acceleration between two parts of his or her body separated by the deviation vector $\xi^{\hat{i}'}$ is given by \cite{morris1}
\begin{equation}
\Delta a^{\hat{j}'}=-c^2 R^{\hat{j}'}_{\; \hat{0}' \hat{k}' \hat{0}'} \xi^{\hat{k}'},
\end{equation}
where $R^{\hat{i}'}_{\; \hat{j}' \hat{k}' \hat{l}'}$ is the Riemann tensor. At the throat, the components of the tidal acceleration are given by
\begin{equation}
\Delta a^{\hat{1}'}\big|_{r_0}=-\frac{\alpha c^2}{r_0^2}\left(1-\frac{2}{3}x\right)\xi^{\hat{1}'}, \hspace{0.2cm} \Delta a^{\hat{2}'}\big|_{r_0}=\frac{1}{r_0^2}\left(1-x\right)\gamma_0^2 v_0^2\xi^{\hat{2}'}
\end{equation}
\begin{equation}
\Delta a^{\hat{3}'}\big|_{r_0}=\frac{1}{r_0^2}\left(1-x\right)\gamma_0^2 v_0^2\xi^{\hat{3}'},
\end{equation}
where $\gamma=\left(1-\frac{v^2}{c^2}\right)^{-\frac{1}{2}}$ and $v=\pm\frac{\sqrt{g_{rr}}dr}{\sqrt{|g_{tt}|}dt}$ is the the radial velocity of the traveler as measured by a static observer. $v_0$ denotes the radial velocity at the throat. Restricting the radial components of the tidal acceleration below one Earth gravity, i.e., $|\Delta a^{\hat{1}'}|\leq g$, we obtain, for a traveler of typical size $|\xi|\sim$ 2 meter,
\begin{equation}
r_0^2\geq \frac{2\alpha c^2}{g}\left(1-\frac{2}{3}x\right).
\end{equation}
For a wormhole with $M>0$, we have $\frac{1}{2}<\alpha \leq 1$ and $0<x<1$. Taking $\alpha \gtrsim \frac{1}{2}$ and $x \lesssim 1$, we obtain a minimum throat radius $r_{0min}=\sqrt{\frac{c^2}{3g}}\simeq 8.64R_E$, where $R_E\simeq 6400$ km is the Earth radius. Thus, the wormhole throat radius must be greater than $8.64$ times the Earth radius. It should be noted that minimum $r_0$ corresponds to minimum $|\kappa|$. We get $|\kappa|_{min}=r_{0min}^2\simeq 3.0\times 10^{15}$ m$^2$. Any $|\kappa|$ or $r_0$ value lying above the corresponding minimum value will satisfy the above constraint for a given $x$ lying in the range $0<x<1$. The authors in \cite{solar} obtain the solar constraint $|\kappa_g|\lesssim 3\times 10^5$ m$^5$ sec$^{-2}$ kg$^{-1}$. In m$^2$ unit, this gives $|k|=\frac{|\kappa_g|}{8\pi G}\lesssim 1.8\times 10^{14}$ m$^2$. The upper limit of the solar constraint is less than $|\kappa|_{min}$. This indicates that, for the solar constraint, the radial components of the tidal acceleration cannot be made less than one Earth gravity. For $|\kappa|=1.8\times 10^{14}$ m$^2$, $|\Delta a^{\hat{1}'}|_{r_0}=\frac{\alpha}{x}\left(1-\frac{2}{3}x\right)\times 10^3$ m sec$^{-2}$ which gives $|\Delta a^{\hat{1}'}|_{min}=\frac{10^3}{6}$ m sec$^{-2}$ $\simeq 17g$. Therefore, for the solar constraint, the minimum radial tidal acceleration at the throat is greater than 17 times the earth gravity. To make this minimum value less than one earth gravity, we must have $|\kappa|>|\kappa|_{min}$. Additionally, the angular components of the tidal acceleration put limits on the radial velocity $v_0$ at the throat, which are reasonable.

\section{Some special cases}
\noindent As a special case, let us first consider $\alpha=1$. We also take $G=1$ and $c=1$. This gives the electrically charged solution discussed in \cite{EiBIcharged1,EiBIcharged2}. The energy density $\rho=\frac{1}{8\pi}\frac{Q^2}{r^4}$ helps us to set $C_0=\frac{Q^2}{8\pi}$ where $Q$ is the charge. In this case, the hypergeometric function can be written in terms of an elliptic function (see Appendix). The properties of this solution have been studied in detail for both positive and negative $\kappa$ \cite{EiBIcharged2,EiBIcharged3,EiBIwormhole2,olmo}. By writing the physical metric in a different gauge and noting the bounce of the radial function, the authors in \cite{EiBIwormhole2} show that  this solution signals the presence of a wormhole for negative $\kappa$. They define a parameter $\delta_1=\frac{Q^{3/2}}{2M|\kappa|^{1/4}}$. They show that if $\delta_1$ is tuned to the value $\delta_1^*=-\frac{1}{\beta}\approx 0.572$, the spacetime becomes regular. In our case also, we get $\delta_1^*=\frac{Q^{3/2}}{2M|\kappa|^{1/4}}\approx 0.572$ [using (\ref{eq:constraint2})] and the charge-to-mass ratio $\frac{Q}{M}\approx \frac{1.144}{\sqrt{x}}$. For $\frac{Q}{M}<1.144$ ($x>1$), the throat at $r=r_0$ is covered by an event horizon. On the other hand, for $\frac{Q}{M}>1.144$ ($x<1$) the horizon is absent and we have a wormhole solution. Therefore, we obtain the critical charge-to-mass ratio $\left(\frac{Q}{M}\right)_c\approx 1.144$.

\begin{figure}[ht]
\centering
\includegraphics[scale=0.85]{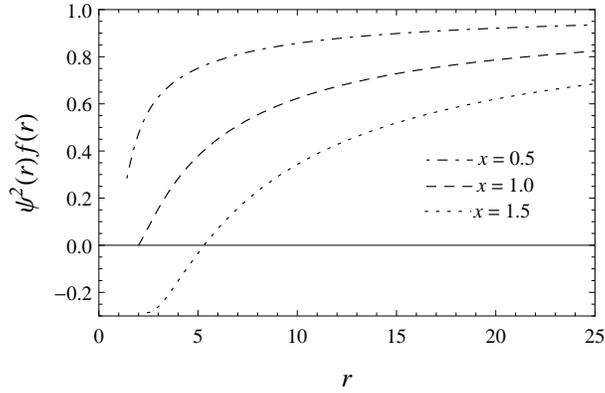}
\caption{Plots of $\psi^2(r)f(r)$ for $\alpha=\frac{3}{4}$ and $\kappa=-4.0$.}
\label{fig:gtt}
\end{figure}

\begin{figure}[ht]
\centering
\includegraphics[scale=0.85]{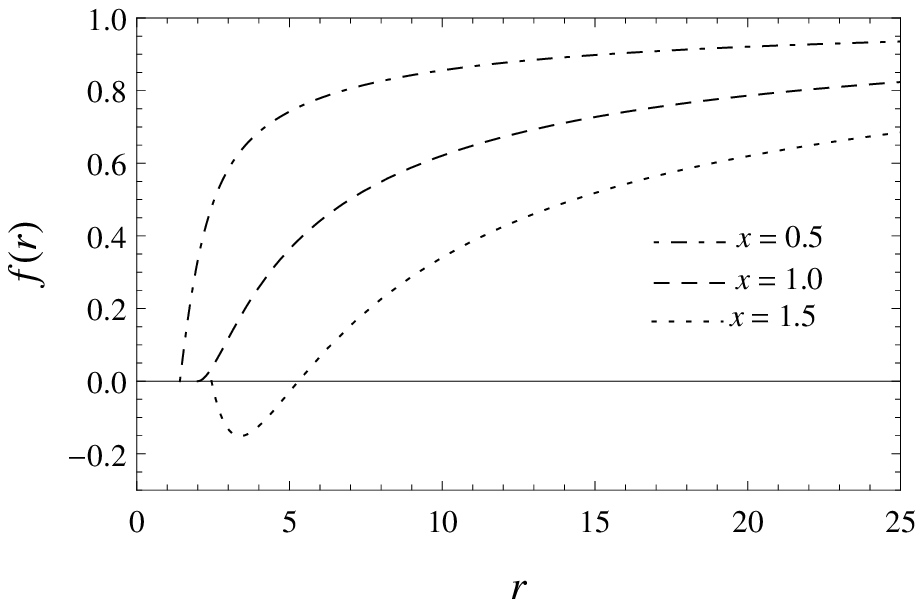}
\caption{Plots of $f(r)$ for $\alpha=\frac{3}{4}$ and $\kappa=-4.0$.}
\label{fig:grr}
\end{figure}

\begin{figure}[ht]
\centering
\includegraphics[scale=0.85]{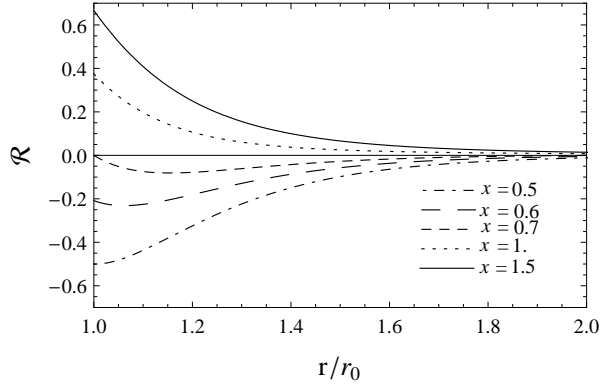}
\caption{Plots of curvature scalar $\mathcal{R}$ for $\alpha=\frac{3}{4}$ and $\kappa=-4.0$.}
\label{fig:curvature}
\end{figure}

\begin{figure}[ht]
\centering
\includegraphics[scale=0.85]{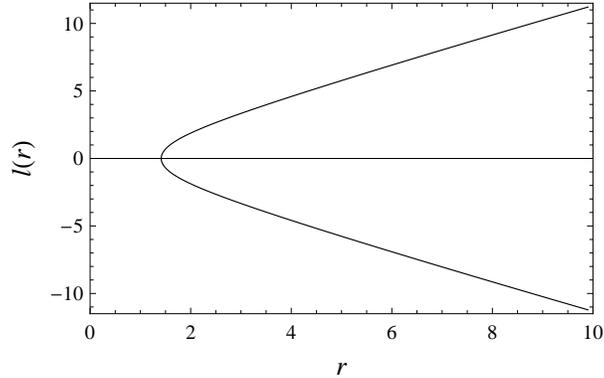}
\caption{Plot of the proper distance $l(r)$ as a function of the radial coordinate $r$ for $\alpha=\frac{3}{4}$, $\kappa=-4.0$ and $x=0.5$.}
\label{fig:proper_distance}
\end{figure}

\begin{figure}[ht]
\centering
\includegraphics[scale=1.0]{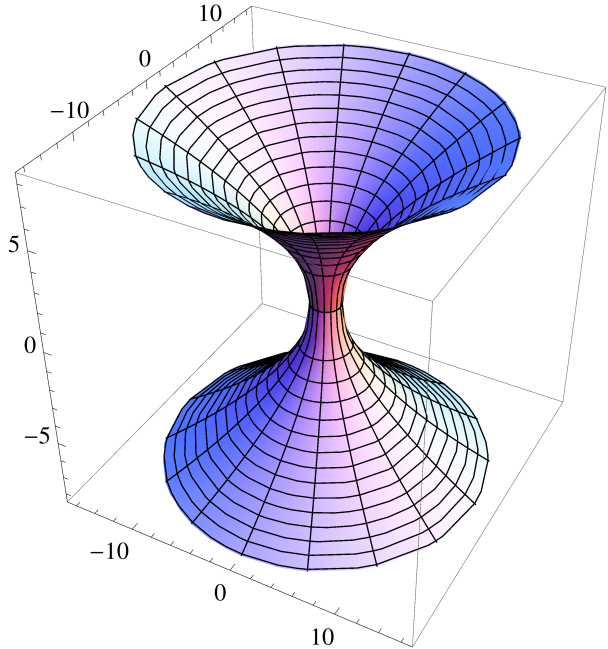}
\caption{Embedding diagram of the wormhole, i.e., the surface of revolution $z(r)$, for $\alpha=\frac{3}{4}$, $\kappa=-4.0$ and $x=0.5$.}
\label{fig:embedding}
\end{figure}

We calculate and plot the metric functions and curvature scalar of the regular spacetime for $\alpha=\frac{3}{4}$, as shown in Figs. \ref{fig:gtt}--\ref{fig:curvature}. It should be noted from both (\ref{eq:curvature}) and the plots, the curvature at $r=r_0$ vanishes for $x=\frac{3\alpha+3}{2\alpha+6}=0.7$ ($\alpha=\frac{3}{4}$). For $0<x\le 0.7$, the curvature remains negative around the throat. For a wormhole, the proper radial distance defined by
\begin{equation}
l(r)=\pm \int_{r_0}^{r} \frac{dr}{\sqrt{f(r)}}
\end{equation}
must be finite at all finite ``$r$" throughout the spacetime. The $\pm$ signs are for the two parts of the wormhole geometry connected by the throat. Figure \ref{fig:proper_distance} shows the dependence of $l(r)$ on $r$ for $\alpha=\frac{3}{4}$. Figure \ref{fig:embedding} shows the embedding diagram of the wormhole for $\alpha=\frac{3}{4}$.



\section{Conclusion}
\noindent In this article, we have exactly solved the field equations in EiBI gravity coupled to an anisotropic fluid having energy momentum tensor of the form $T^\mu_{\;\nu}=diag(-\rho,p_r,p_\theta,p_\theta)=diag(-\rho,-\rho,\alpha\rho,\alpha\rho)$. For this energy momentum tensor, we have shown that, if the matter has to satisfy the energy conditions, the Eddington-Born-Infeld theory parameter $\kappa$ must be negative to have a wormhole geometry. We have obtained the important relation (\ref{eq:constraint2}), between $|\kappa|$, Schwarzschild mass $M$ and the integration constant $C_0$ (related to the energy density) to make the spacetime regular at the throat and, hence, to make the wormhole. We have also obtained a restriction on $\alpha$ to have a wormhole geometry. We have defined a parameter $x=\frac{r_0^2}{|\kappa|}$. For $x<1$, we have a wormhole geometry. But, for $x>1$, an event horizon is formed around the throat giving a regular black hole with singularity removed by a wormhole throat. We have obtained constraints on both $|\kappa|$ and the wormhole throat to restrict the tidal acceleration (at the throat) below one Earth gravity. As a special case ($\alpha=1$), we obtained the wormhole supported by an electric field \cite{EiBIwormhole2}. In this case, the critical value $x_c=1$ gives a critical charge-to-mass ratio $\left(\frac{Q}{M}\right)_c\approx 1.144$. Above this critical value, we have a wormhole geometry. But, below this critical value the wormhole throat is covered by an event horizon. The condition in (\ref{eq:constraint1}) which leads to $\delta_1^*\approx 0.572$ for $\kappa<0$ and $\alpha=1$ makes the spacetime nonsingular. Therefore, for $\delta_1^*\neq 0.572$, we have either singular black hole or wormhole with singular throat, depending on the values of the parameters. Hence, any attempt to break the tuning of $\delta_1^*\approx 0.572$, will turn the regular wormhole to either a singular black hole or nontraversable wormhole.

The general solution for $\kappa>0$ gives a singular solution. One can use this singular spacetime as well as the wormhole and the regular black hole spacetime to study various aspects such as geodesic structure, gravitational lensing, particle collisions, quasinormal modes, etc. Various properties such as geodesic completeness, behavior of timelike congruence and scattering of waves in the wormhole spacetime supported by the electric field have been studied recently \cite{olmo}. They show that, irrespective of the presence of the curvature divergence, the geodesics are complete, timelike geodesic congruences do not experience any pathological behavior, and a scalar wave propagating in the background is well behaved throughout the spacetime. It would be interesting to know whether these properties differ for the $\alpha\neq 1$ general spacetimes discussed in this paper. In our case, the radial pressure is negative. However, one can consider a more realistic energy momentum tensor and try to obtain an exact wormhole geometry with matter satisfying all the energy conditions. It would be useful to know if there exists a wormhole solution for $\kappa>0$, with stress energy satisfying all the energy conditions.

\section*{Acknowledgments}
\noindent The author acknowledges the Council of Scientific and Industrial Research, India, for providing support through a fellowship. He also acknowledges Sayan Kar for going through the manuscript and making suggestions which helped to improve the manuscript. He also thanks the anonymous referee for making critical comments which helped to improve the manuscript.

\section{Appendix: Useful integrations}
The integration in (\ref{eq:integration}) can be evaluated as follows:
\begin{eqnarray*}
\int\frac{H^2H'}{1+\frac{\bar{\kappa} C_0}{r^{2(\alpha+1)}}}dr &=& \int r^2\frac{d}{dr}\left(\sqrt{r^2+\frac{\bar{\kappa} C_0}{r^{2\alpha}}} \right)dr\\
&=& r^3\sqrt{1+\frac{\bar{\kappa} C_0}{r^{2(\alpha+1)}}}-2\int r^2\sqrt{1+\frac{\bar{\kappa} C_0}{r^{2(\alpha+1)}}}dr\\
&=& \frac{r^3}{3}\sqrt{1+\frac{\bar{\kappa} C_0}{r^{2(\alpha+1)}}}-\frac{2}{3}(\alpha+1)\bar{\kappa} C_0 I(r),
\end{eqnarray*}
where
\begin{equation*}
I(r)=\int \frac{1}{r^{2\alpha}\sqrt{1+\frac{\bar{\kappa} C_0}{r^{2(\alpha+1)}}}} dr.
\end{equation*}
Performing the above integration, we obtain
\begin{equation}
I = \left\{
  \begin{array}{lr}
    \frac{2}{3}\log\left[\left(\frac{r}{r_0}\right)^{\frac{3}{2}}+\sqrt{\left(\frac{r}{r_0}\right)^{3}\mp 1}\right] & : \alpha =\frac{1}{2}\\
  \frac{r^{1-2\alpha}}{1-2\alpha} {}_2F_1\left[\frac{1}{2},\frac{2\alpha-1}{2\alpha+2},\frac{4\alpha+1}{2\alpha+2};\pm\left(\frac{r_0}{r}\right)^{2\alpha+2} \right] & : \alpha \neq \frac{1}{2}
  \end{array}
\right. ,
\nonumber
\end{equation}
where upper and lower signs are for $\kappa<0$ and $\kappa>0$, respectively, and $r_0=(|\bar{\kappa}| C_0)^{\frac{1}{2(\alpha+1)}}$. For $\alpha=1$ and $\kappa<0$, the hypergeometric function can be written in terms of the elliptic function. This can be done by putting $r_0/r=\sin y$ in the last integral:
\begin{eqnarray*}
I(r) &=& -\frac{1}{r_0}\int \frac{1}{\sqrt{1+\sin^2y}} dy\\
&=& -\frac{1}{r_0}EllipticF\left[\sin^{-1}\left(\frac{r_0}{r}\right),-1\right]
\end{eqnarray*}

\end{document}